\documentclass[]{iau}

\usepackage{amsmath}
\usepackage{graphicx}
\usepackage{xcolor}
\usepackage{multirow}
\usepackage{csquotes}

\begin{document}

\lefttitle{F.~Eppel et al.}
\righttitle{VLBI Scrutiny of a New Neutrino-Blazar Multiwavelength-Flare Coincidence}

\jnlPage{1}{7}
\jnlDoiYr{2021}
\doival{10.1017/xxxxx}
\aopheadtitle{Proceedings IAU Symposium}
\volno{375}
\editors{Y.\,Liodakis eds.}

\title{VLBI Scrutiny of a New Neutrino-Blazar Multiwavelength-Flare Coincidence}

\author{F.~Eppel$^1$, M.~Kadler$^1$, E.~Ros$^2$, F.~Rösch$^1$, J.~Heßdörfer$^1$, P.~Benke$^2$, P.~G.~Edwards$^3$, C.~M.~Fromm$^1$, M.~Giroletti$^4$, A.~Gokus$^{5}$, J.~L.~G\'omez$^6$, S.~Hämmerich$^{7}$, D.~Kirchner$^1$, Y.~Y.~Kovalev$^{2,8}$, T.~P.~Krichbaum$^2$, M.~L.~Lister$^9$, C.~Nanci$^4$, R.~Ojha$^{10}$, G.~F.~Paraschos$^2$, A.~Plavin$^8$, A.~C.~S.~Readhead$^{11}$, J.~Stevens$^3$, P.~Weber$^1$}
\affiliation{$^1$\,Julius-Maximilians-Universität Würzburg, Physikalisches Institut, Lehrstuhl für Astronomie, Emil-Fischer-Straße 31, 97074 Würzburg, Germany -- $^2$\,Max-Planck-Institut für Radioastronomie, Auf dem Hügel 69, 53121 Bonn, Germany -- $^3$\,CSIRO Astronomy and Space Science, ATNF, PO Box 76. Epping NSW 1710. Australia -- $^4$\,INAF, Istituto di Radio Astronomia, Via Piero Gobetti, 41029 Bologna, Italy -- $^{5}$~McDonnell Center for Space Science, Washington University, MSC 1105-109-03, One Brookings Drive, St. Louis, MO 63130-4899, USA --  $^6$\,Instituto de Astrof\'{\i}sica de Andaluc\'{\i}a, Gta. de la Astronomia, 18008 Granada, Spain -- $^{7}$ Remeis Observatory and Erlangen Centre for Astroparticle Physics, Universität Erlangen-Nürnberg, Sternwartstr. 7, 96049 Bamberg, Germany --  $^8$\,Lebedev Physical Institute, Leninskii pr. 53, 119991, Moscow, Russia -- $^9$\,Department of Physics and Astronomy, Purdue University, 525 Northwestern Avenue, West Lafayette, IN 47907, USA -- $^{10}$\,NASA, Goddard Space Flight Center, Astrophysics Science Division, Code 661, Greenbelt, MD 20071, USA -- $^{11}$\,California Institute of Technology, 1200 E California Blvd, Pasadena, CA 91125, USA}

\begin{abstract}
In the past years, evidence has started piling up that some high-energy cosmic neutrinos can be associated with blazars in flaring states. On February 26, 2022, a new blazar-neutrino coincidence has been reported: the track-like neutrino event IC220225A detected by IceCube is spatially coincident with the flat-spectrum radio quasar PKS\,0215+015. Like previous associations, this source was found to be in a high optical and $\gamma$-ray state. Moreover, the source showed a bright radio outburst, which substantially increases the probability of a true physical association. We have performed six observations with the VLBA shortly after the neutrino event with a monthly cadence and are monitoring the source with the Effelsberg 100m-Telescope, and with the Australia Compact Telescope Array. Here, we present first results on the contemporary parsec-scale jet structure of PKS\,0215+015 in total intensity and polarization to constrain possible physical processes leading to neutrino emission in blazars.
\end{abstract}


\maketitle
\section{Introduction}
Gamma-ray blazars have long been predicted to be bright enough in neutrino emission and to exhibit a neutrino spectrum flat enough to exceed the atmospheric background at sufficiently high energies above about 100\,TeV \citep[e.g.,][]{Mannheim1992}. 
The fundamental assumption in such calculations is that both neutrinos and $\gamma$-ray photons are produced by the same accelerated protons in the jet (hadronic processes as the dominant origin for the $\gamma$-ray emission of blazars).  
\cite{Kadler2016} have shown that within this general framework the long-term average $\gamma$-ray emission of blazars as a class is indeed in agreement with both the measured all-sky flux of very-high-energy neutrinos, and the spectral slope of the IceCube signal. 
However, the association of an individual blazar with a single neutrino has been limited by the poor positional accuracy of the shower-like events considered in that study.
A major step forward was made when the track-like muon neutrino event IC\,170922A was found to coincide with the $\gamma$-ray blazar TXS\,0506+056 (GCN Circular 21916).
Thanks to the much improved positional accuracy of the reconstruction for that event and the variability of the $\gamma$-ray emission, the association between IC\,170922A and TXS\,0506+056 was found to be on the level of 3\,$\sigma$ \citep{IceCube2018}.
The general statistical correlation between radio-bright AGN samples and IceCube neutrinos that was revealed only recently \citep{Plavin2020,Plavin2021,Hovatta2021}
is a fundamental independent observational breakthrough that
strongly calls for high angular-resolution radio data of individual coincident neutrino detections and radio flares to solve this important question \citep[e.g.,][]{Nanci2022}. On February 26, 2022, the IceCube collaboration announced a new high-energy neutrino alert (IC220225A\footnote{\url{https://gcn.gsfc.nasa.gov/gcn/gcn3/31650.gcn3}}) which is spatially coincident with the flat-spectrum radio quasar (FSRQ) PKS\,0215+015. Like previous neutrino-blazar associations this source was found to be in a high-optical and $\gamma$-ray state \citep{FermiATel,NesciAtel}. On top of that, members of our team found the source to be in a bright radio outburst \citep{KadlerAtel,PlavinAtel}, which increases the probability of a true physical association. In the following we discuss the first results of our ongoing follow-up observing campaign of the event with the ATCA, the Effelsberg 100-m telescope and the VLBA.

\begin{figure}
\centering
\includegraphics[width=.9\columnwidth]{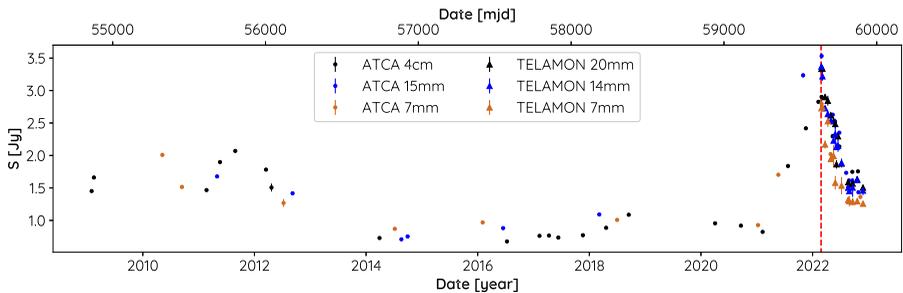}
\caption{Radio Light Curve of PKS\,0215+015 from TELAMON (Effelsberg) and ATCA observations since 2009. The source exhibits a major radio flare coincident with neutrino alert IC220225A (dashed red line). This radio flare is also coincident with a $\gamma$-ray flare observed by \textit{FERMI-LAT}}
\label{atca-telamon-lc}
\end{figure}

\section{Observations \& Analysis}
ATCA has monitored PKS\,0215+015 at multiple frequencies ranging from 2.1\,GHz to 50\,GHz since 2004 as part of calibrator and AGN monitoring programs \citep{Stevens2012}. The data are being processed by a standard pipeline and are made publicly available\footnote{\url{https://www.narrabri.atnf.csiro.au/calibrators/calibrator_database_viewcal.html?source=0215+015&detailed=true}}. Moreover, we included the source into the regular monitoring sample of the TELAMON program \citep{TELAMON} shortly after the neutrino detection (first observation on Feb 27, 2022), which is being carried out with the Effelsberg 100-m telescope. TELAMON flux densities are derived by comparison with suitable secondary calibrators and averaged over three different sub-bands (20\,mm, 14\,mm and 7\,mm), a detailed description will be published elsewhere. 
Triggered by the intriguing association of the neutrino event with the flare,
 we have succesfully requested VLBA target-of-opportunity observations of PKS\,0215+015 for six different sessions distributed over five months after the neutrino event ($\sim$\,one per month) through a director's discretional time (DDT) proposal. These observations were carried out at 2\,cm, 1.3\,cm and 7\,mm including full polarization information for every band. In addition to this DDT request the source has been observed with the VLBA at 2\,cm as part of the MOJAVE program in the past \citep{Lister2019} and was reinstated into the regular monitoring just shortly before the neutrino event. 

\begin{figure}
\includegraphics[width=0.48\columnwidth]{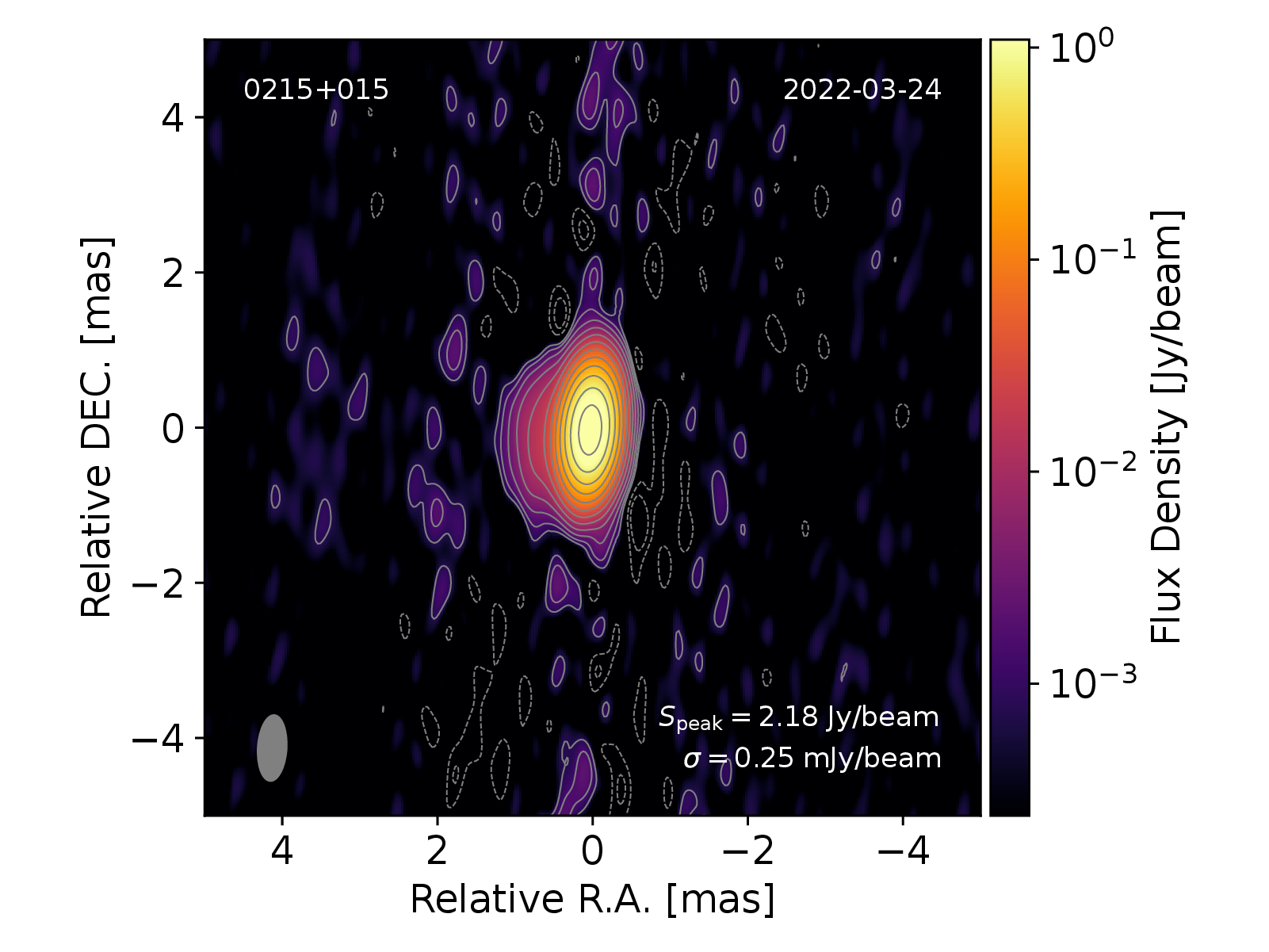}
\includegraphics[width=0.48\columnwidth]{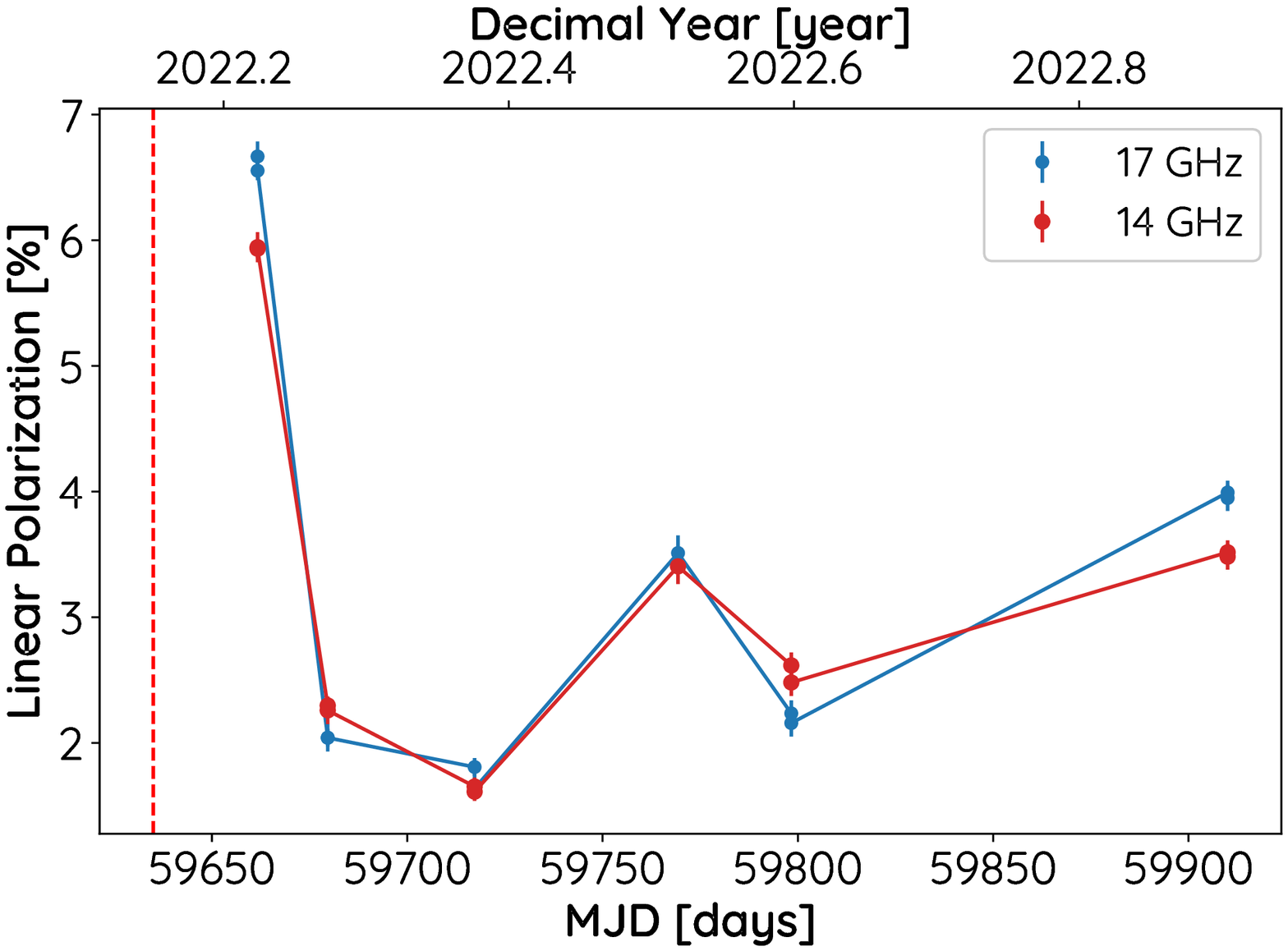}
\caption{\textit{Left:} VLBI-Image of PKS\,0215+015  in total intensity at a wavelength of 1.3\,cm (23\,GHz) observed with the VLBA on March 24, 2022, one month after the neutrino-detection. \textit{Right:} Linear polarization fraction of the same source as observed with the Effelsberg 100-m telescope as part of the TELAMON program. The dashed red line indicates the time of the neutrino event.}
\label{plot2}
\end{figure}

\section{First Results}
\subsection{Radio-Lightcurve}
Figure\,\ref{atca-telamon-lc} shows the radio light curve of PKS\,0215+015 compiled from selected ATCA data since 2009 and TELAMON observations. The source clearly reached its historical maximum near the time of the neutrino event and has been decreasing in radio brightness since. Publicly available \textit{Fermi-LAT} data indicate a simultaneous $\gamma$-ray outburst\footnote{\url{https://fermi.gsfc.nasa.gov/ssc/data/access/lat/LightCurveRepository/source.php?source_name=4FGL_J0217.8+0144}}.

\subsection{VLBI-Structure}
The VLBA data were processed in a standard way using AIPS \citep{AIPS} for delay and amplitude calibration and DIFMAP \citep{DIFMAP} for self-calibration and imaging. Here, we focus on the data from the first observing epoch on March 24, 2022. Figure\,\ref{plot2} (left) shows the first image of PKS\,0215+015 after the neutrino event at 23\,GHz from this epoch. The source shows a very compact structure with a jet pointed in the Eastern direction with one visible jet component. This jet component is also visible in the 2\,cm and 7\,mm images from March 24, 2022, and partially blended with the core in all three wavebands. We have estimated the brightness temperature $T_B$ of the core component using the method described by \cite{Kovalev2005}. At 2\,cm and 1.3\,cm, we find $T_B \sim 2.2 \cdot 10^{13}$\,K, while at 7\,mm $T_B \sim 3.5 \cdot 10^{12}$\,K. We note that all derived brightness temperatures exceed the equipartition limit \citep{Readhead1994} and therefore suggest Doppler beaming.

\subsection{Polarization Properties}
From the Effelsberg observations, we derived polarization properties, namely linear polarization and electric vector position angle (EVPA) of the radio emission. 
Figure\,\ref{plot2} (right) shows preliminary results of this analysis, indicating that the source was in an elevated polarization state close to the neutrino event ($\sim6$\,\% linear polarization at 2\,cm) which decayed afterwards. 
The EVPA does not show any significant changes. Moreover, we have performed a preliminary polarization calibration of the first VLBA epoch using \textsc{PolSolve} \citep{Ivan} in \textsc{CASA} which suggests that the polarized emission is core-dominated, just like the emission in total intensity. 

\section{Conclusions \& Outlook}

Just like the three blazars PKS\,1424-418, TXS\,0506+056 and PKS\,1502+106, which were previously associated with neutrino events \citep{Kadler2016,Ros2020,Karamanavis2015}, PKS\,0215+015 exhibits a very compact core-jet morphology . This suggests that neutrino production in PKS\,0215+015 happens close to the central core. In the case of TXS\,0506+056 and PKS\,1424-418, the neutrino-associated radio flare also happened mostly in the compact core region. In TXS\,0506+056, the core flux density increase was accompanied by a core expansion with apparent superluminal velocity \citep{Ros2020}. Moreover, a limb-brightened structure, as predicted by \cite{Tavecchio2015}, was found in TXS\,0506+056, PKS\,1424-418 and PKS\,1502+106. The similar sub-parsec scale structures of these possible neutrino-blazar associations could therefore indicate that neutrino-production in blazars usually happens in similar environments and scales. In order to further investigate this, we will search the remaining five VLBA epochs of PKS\,0215+015 for these signatures. On top of that, we will investigate the kinematics of the jet component to find out whether its ejection happened close in time to the neutrino event. Due to its compactness it can be challenging to detect a limb-brightening in the jet, however the collected polarization data (i.e., the orientation of the EVPA) can equally provide evidence for the existence of a \enquote{spine-sheath} structure.

\section*{Acknowledgements}
This work is based on observations with the 100-m telescope of the MPIfR (Max-Planck-Institut für Radioastronomie) at Effelsberg. FE, FR, JH, and MK acknowledge support from the Deutsche Forschungsgemeinschaft (DFG, grants 447572188, 434448349, 465409577). 
We acknowledge the M2FINDERS project from the European Research Council (ERC) under the European Union’s Horizon 2020 research and innovation programme (grant agreement No 101018682).
This research has made use of data from the MOJAVE database that is maintained by the MOJAVE team \citep{Lister2019}. The National Radio Astronomy Observatory is a facility of the National Science Foundation operated under cooperative agreement by Associated Universities, Inc.

\bibliographystyle{}

\end{document}